\documentclass{article}
\usepackage[english]{babel}
\usepackage[utf8]{inputenc}
\usepackage{johd}
\usepackage{amsmath}
\usepackage{caption}
\usepackage{subcaption}

\title{Turing Machines Equipped with CTC in Physical Universes}

\author{Sara Babaee K.$^{1}$$^{*}$, Farzad Didehvar$^{1}$$^{\dagger}$ \\
        \small $^{1}$Department of Mathematics and Computer Science, Amirkabir University of Technology, Tehran, Iran \\
        \\
        \small $^{*}$Email: \tt{sarababaei@aut.ac.ir} \\
        \small $^{\dagger}$Email: \tt{didehvar@aut.ac.ir} \\
}

\date{} 

\begin{document}

\maketitle

\begin{abstract} 
We study the paradoxical aspects of closed time-like curves and their impact on the theory of computation. After introducing the $\text{TM}_\text{CTC}$, a classical Turing machine benefiting CTCs for backward time travel, \textit{Aaronson} et al. proved that $\text{P} = \text{PSPACE}$ and the $\Delta_2$ sets, such as the halting problem, are computable within this computational model.
\\
Our critical view is the physical consistency of this model, which leads to proposing the \textit{strong axiom}, explaining that every particle rounding on a CTC will be destroyed before returning to its starting time, and the \textit{weak axiom}, describing the same notion, particularly for Turing machines. We claim that in a universe containing CTCs, the two axioms must be true; otherwise, there will be an infinite number of any particle rounding on a CTC in the universe.
\\
An immediate result of the \textit{weak axiom} is the incapability of Turing machines to convey information for a full round on a CTC, leading to the proposed $\text{TM}_\text{CTC}$ programs for the aforementioned corollaries failing to function. We suggest our solution for this problem as the \textit{data transferring hypothesis}, which applies another $\text{TM}_\text{CTC}$ as a means for storing data. A prerequisite for it is the existence of the concept of Turing machines throughout time, which makes it appear infeasible in our universe. Then, we discuss possible physical conditions that can be held for a universe containing CTCs and conclude that if returning to an approximately equivalent universe by a CTC was conceivable, the above corollaries would be valid.
\end{abstract}

\noindent\\

\section{Introduction}

Roughly speaking, \textbf{space-time} is a four-dimensional continuous coordinate system that combines the three spatial dimensions (i.e., Euclidean space) with one-dimensional time in such a way that the four axes are not independent. For instance, time passes slower in higher velocities according to special relativity. \citep{einstein-special-relativity}
\\
Thus, space-time consists of points, which we name \textbf{events}, that can be demonstrated as $(x, y, z, t)$ and are used to show the coordinates of particles of the universe. Also, the path that is taken by a particle in space-time is called its \textbf{world line}.

Physical phenomena can be explained in variant types of space-times, such as flat, like Minkowski, or curved space-times. In each of them, numerous properties may appear, such as chronology-violating, meaning that an event $a$ preceding an event $b$ might occur after $b$. Various unfamiliar and somehow unusual features, like wormholes and singularities, tend to emerge in space-time owning this property. \citep{deutsch-ctc} 
An example of the chronology-violating space-times we want to study in this paper is a universe containing CTCs, which entails curved space-time.

To realize what exactly a CTC is, we should first review some concepts. Let us consider space-time as a three-dimensional coordinate system in which a plane graded in the $ct$\footnote{$c$ is the universal constant for speed of light in vacuum.} unit represents space, and the orthogonal axis on the plane graded in the $t$ unit displays the time.
Now, if we assume a ray of light on the origin of space-time, since it moves with velocity $c$, in the next time unit, $t$, it will have taken a $ct$ distance. Thus, in time $t$, the light will be at any point on the circle with origin $(0, 0, t)$ and radius $ct$.
Due to the fact that time and space are continuous, the possible paths the light ray can take illustrate a cone with an apex at the origin called the \textbf{light cone}. This cone states the \textbf{future} or positive direction of time. Similarly, the possible paths that light has taken to reach the origin of space-time also form another cone, which expresses the \textbf{past} or negative direction of time. 

Any vector $v = (x, y, z, t) = (\vec{d}, t)$\footnote{Let $\vec{d} = (x, y, z)$.} in space-time, demonstrates a movement between two points $A$ and $B$ with spatial difference $\Vec{d}$ in time $t$ and based on the relation between $\Vec{d}$ and $t$ is divided into three categories. 

If $|\Vec{d}| < ct$, the path $\Vec{d}$ has been taken with a velocity less than light's and so can be in a particle's world line. In this case, the vector is named \textbf{time-like} and is inside the light cone.

If $|\Vec{d}| = ct$, the path $\Vec{d}$ has been taken with the velocity of light and so cannot be in any particle's world line. In this case, the vector is named \textbf{light-like} and is on the light cone.

Finally, if $|\Vec{d}| > ct$, the path $\Vec{d}$ has been taken with a velocity more than light's and so cannot be in any particle's or light's world line. In this case, the vector is named \textbf{space-like} and is outside the light cone.

A \textbf{Closed Time-like Curve}, or \textbf{CTC} for short, is a time-like line, in the above definition, with the same starting and ending points. Since it is time-like, can be a particle's world line. Any particle of the system owning this kind of world line may return to a state of its past, or in other words, a coordinate of space and time that has been before.

The possibility of the existence of CTCs was raised for the first time by a dutch mathematician, \textit{Willem Jacob van Stockum}. \citep{van-ctc} After that, \textit{Kurt Gödel} introduced the Gödel metric as a solution to Einstein’s field equations, leading to express a universe containing CTCs. \citep{godel-ctc} Factually, \textit{Einstein} formulated the \textbf{field equations}, also known as \textbf{EFE}, within the general theory of relativity in \citep{einstein-general-relativity}. Since then, several solutions have been found for EFE, which are called \textbf{metrics}. These solutions tend to describe universes that include exotic features; for instance, black holes in the Schwarzschild metric, traversable wormholes in the Morris–Thorne metric, and CTCs in the Gödel metric.

Intuitively, a space-time equipped with CTCs provides the possibility of traveling back in time. Similar to someone starting going rightward on the spherical earth and finally reaching a coordinate lefter than their departure point, a particle of the system can enter the CTC and move forward on it and, by the passage of time, since CTC is closed, eventually arrives at a time before its travel’s starting. \citep{sep-time-travel}

\subsection{Paradoxes Come with CTC}

At first sight, time travel causes several logical as well as intuitional paradoxes. In this section, we study the definition, scenario, and possibly the given responses to some sorts of these paradoxes. Then, we discuss a similar issue in section 2 and proceed with trying to respond to it.

\subsubsection{Consistency Paradoxes}
A prominent sort of time travel paradox is the \textbf{consistency paradox} which happens by performing some changes to the past. A convenient example of that is the \textit{grandfather paradox}, which can be described as follow:

\begin{quote}
    Imagine a person who travels back in time to one of their ancestors' eras and kills their grandfather. As a result, they will never be born, so they will not have time-traveled and killed their grandfather. Consequently, they are born and travel to the past, and so on. \citep{aaronson-complexity}
\end{quote}

There also exist a few other equivalent scenarios for the grandfather paradox, such as the story that \textit{Hawking} and \textit{Ellis} claim:

\begin{quote}
    Suppose that with a suitable rocket ship, a person travels in time to arrive before their departure. They can alter any past events only if we assume they have \textit{free will},  leading to stopping themselves from setting out on their travel. Consequently, they will not travel in time, and therefore, nothing will happen to prevent their time travel. \citep{hawking-grandfather}
\end{quote}
Or the \textit{Hitler Paradox}:

\begin{quote}
    The first thing that may cross the minds of some humanitarians to do with the ability of time travel is going back to some time before 1939 and killing Adolf Hitler to prevent the wage of World War II. However, his murder, aside from the widespread impact that might have on the future, wipes the reason\footnote{Unlike the other mentioned scenarios, in the \textit{Hitler Paradox}, instead of physically preventing time travel, the reason for it vanishes.} for its own happening. Meaning that, without the tragedy of WWII, there is no motivation for the time traveler to kill Hitler. \citep{brennan-hitler}
\end{quote}

Basically, during all of these equivalent scenarios, an event $a$ in the past \textbf{influences} an event $b$ in the future, for example traveling in time to the past, which then causes the occurrence of the event $a$.

\subsubsection{Fermi Paradox}
One other type of contradiction is similar to the \textbf{Fermi paradox}, which can be stated as \textit{if traveling back in time was possible, where would future time travelers be right now?} \citep{jones-fermi}
\\
Various answers are given to this paradox, such as time travel may be extremely expensive or dangerous \citep{sep-time-travel} or reaching our era could be impossible by traveling in time for space-time might not be warped enough in our time to allow the existence of closed time-like lines. \citep{hawking-time-travel}

\subsubsection{Newcomb Paradox}
A further contradiction that can be referred to regarding time travel is known as \textbf{Newcomb’s paradox}, in which a game is played as follows: \citep{wolpert-newcomb}

\begin{itemize}
    \item Two players in the roles of “predictor” and “chooser” play the game with two boxes, namely $A$, containing 1000\$, and $B$, which is empty.
    
    \item The \textit{chooser} will select either both boxes $A$ and $B$ or just box $B$ in their turn and will be paid all the money in their chosen box(es).

    \item The \textit{predictor} will do one of the following in their turn:

    \begin{itemize}
        \item If they foresee that the other player would select both boxes, they will leave box $B$ empty.

        \item Otherwise, if they predict that the other player would select just box $B$, they will put 1’000’000\$ in it.
    \end{itemize}
\end{itemize}

\begin{figure}[H]
    \centering
    \includegraphics[width=0.3\textwidth]{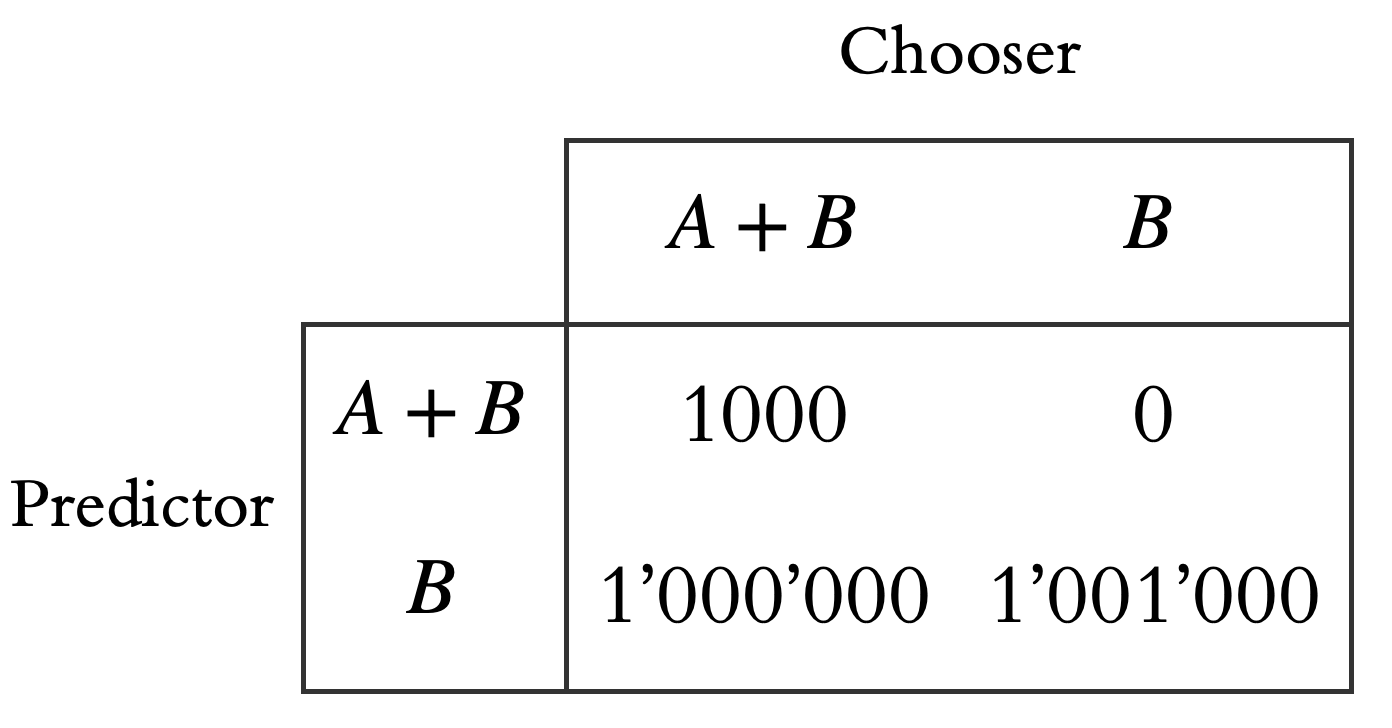}
    \caption{\label{fig1}Newcomb’s game payoff matrix.}
\end{figure}

In game theory, based on the \textit{strategic dominance principle}, the chooser should always choose both boxes $A$ and $B$, whereas using the \textit{expected utility principle} and the fact that the predictor is “infallible”, the chooser should always take box $B$. \citep{nozick-newcomb}

In philosophy, it is believed that perfect prediction or time travel, which can be used as a tool for perfect prediction, conflicts with \textit{free will}; for the sake that it is not recognizable that the prediction is the result of choice or vice versa. \citep{craig-newcomb}

\subsubsection{Causality Loop}
A \textbf{causality loop} is another paradox that might occur via time travel. It means that an event $a$ in the past \textbf{causes}\footnote{It is worth noticing the difference between the used terms \textbf{influence} and \textbf{cause} when defining the consistency paradoxes and the causality loop, respectively.} an event $b$ in the future, which is indeed the cause of $a$. Then, both events exist in space-time, while their origin cannot be determined. \citep{lobo-causality}

\subsubsection{Knowledge Creation Paradox}
The paradox that might happen by traveling in time and does not vanish by consistency methods since it is not known as a logical contradiction at all is the \textbf{knowledge creation paradox}. To comprehend it, suppose that someone travels back in time to reach Gödel’s era and meets him before 1931, the publishing date of his incompleteness theorem paper, where they dictate Gödel the paper. As a result, he admires them and publishes the paper as expected. Thus, it is said that every occurrence in the world, with and without their time travel, is identical, and nothing paradoxical happens, excluding that neither Gödel nor the time traveler genuinely thought about and produced the contents of the paper. In other words, there is no original point of creation for the incompleteness theorem, and knowledge has been created without anyone putting effort into it.

This non-intuitive feature of time travel, which is thought to be preserved in CTCs, is the foundation of related results in \citep{aaronson-complexity} and \citep{aaronson-computability}. They try to solve a hard problem without allocating the desired amount of time or memory to it.

However, we argue that by considering a more precise formulation of the scenario, the universe is not entirely the same in both visits from an outward observer's point of view; in the first, Gödel thinks about the incompleteness theorem, while in the second visit, he communicates with a time traveler. Therefore, this scenario also can be logically paradoxical.
\\
Also, the aforementioned scenario can be seen as an instance of causality loops since it is not recognizable that the time traveler induced Gödel the incompleteness theorem or vice versa; they learned it from Gödel.
\\

Here, we should declare that in a universe containing CTCs, the whole universe, including all creatures, indeliberately return to a time coordinate resulting in the universe being identical in any visit. In contrast, through time travel, just an individual travels to a specific moment. For more explanation, considering the story of the \textit{grandfather paradox}, let us see the problem from the viewpoint of an observer out of space-time who does not move on a CTC. Then for them, the world is not exactly as it was after the grandson's time travel, since in the first view of the coordinate of space-time, the grandson does not exist; however, in the second view, he stands alongside his grandfather.
Hence, arriving at an already-been moment via CTC is not equivalent to an individual’s time travel.
\\
To our best knowledge, the difference between these two concepts has not been discussed sufficiently, which persuaded us to think about potential problems that might arise from CTCs.

\subsection{CTC Consistency}

So far, various people with different approaches have tried to create conditions to make CTCs consistent and eliminate their associated paradoxes. For instance, \textbf{Novikov’s self-consistency principle} explains that events that alter the past occur with a probability equal to zero. Therefore, by excluding all the self-inconsistent happenings, time travel paradoxes vanish. \citep{novikov-consistency} Here, we explain the method that \textit{Deutsch} proposed in \citep{deutsch-ctc} and work based on it.

In \citep{deutsch-ctc}, he studied the physical effects of the existence of CTCs with a quantum computational approach and modeled the computations using CTCs. Furthermore, assuming that classical physics has a minimum approximate consistency near closed time-like curves, he showed that chronology violation might place conflicting constraints on inputs that result in paradoxes in the classical case. Even though, in the quantum case, all these paradoxes are avoidable.

At first, \textit{Deutsch} generally claims that every computational network in which there is a closed path in space-time for the information can be converted into a simplified standard computational network, which with every set of inputs, generates the same outputs as the original network. Also, in this equivalent network, bits interact only in gates, where operations are performed, too. The $n$ bits on closed time-like paths first go to an ambiguous future for all gates. Then go back to the past of all gates with a negative delay and, finally, resume their original paths. If the required time for passing all gates in the computational network was $T$, the bits delay time, or equivalently the time each bit travels to the past, is $-T$. At the same time, $m$ bits from a definite past enter the network as input, creating the output by communication with $n$ CTC bits, and going to an unambiguous future.

In addition, \textit{Deutsch} expresses that for CTCs to be consistent, the evolutionary operator of each gate must have a fixed point. Hence, this point would be the stable state of the information on closed time-like paths. He also shows that such a fixed point always exists in the quantum case, although it may not exist in the classical case.

\subsection{$\text{TM}_{\text{CTC}}$ Computational Model}

In 2016, \textit{Scott Aaronson}, based on Deutschian CTC, proposed the Turing machine equipped with CTC, named $\text{TM}_{\text{CTC}}$, in the classical and $\text{QTM}_{\text{CTC}}$ in the quantum case. In the following, we will discuss the classical computational model.
\\\
The $\text{TM}_{\text{CTC}}$ has two different types of memory registers (Turing machine’s tapes):

\begin{quote}
    $\textbf{R}_{\textbf{CR}}$: Registers that respect the chronology, coming from a known past and going to an unambiguous future. Note that the input of the model will be on these registers.
    \\\
    $\textbf{R}_{\textbf{CTC}}$: Registers round on the CTC.
\end{quote}

\begin{figure}[H]
    \centering
     \begin{subfigure}[b]{0.45\textwidth}
         \centering
         \includegraphics[width=\textwidth]{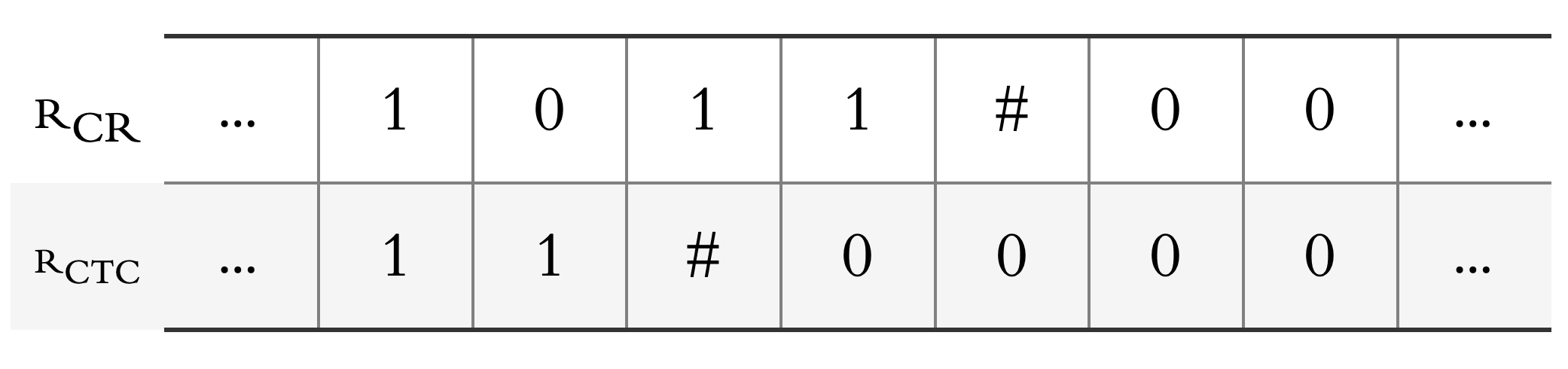}
         \caption{Tape}
         \label{fig2-a}
     \end{subfigure}
     \hfill
     \begin{subfigure}[b]{0.45\textwidth}
         \centering
         \includegraphics[width=0.65\textwidth]{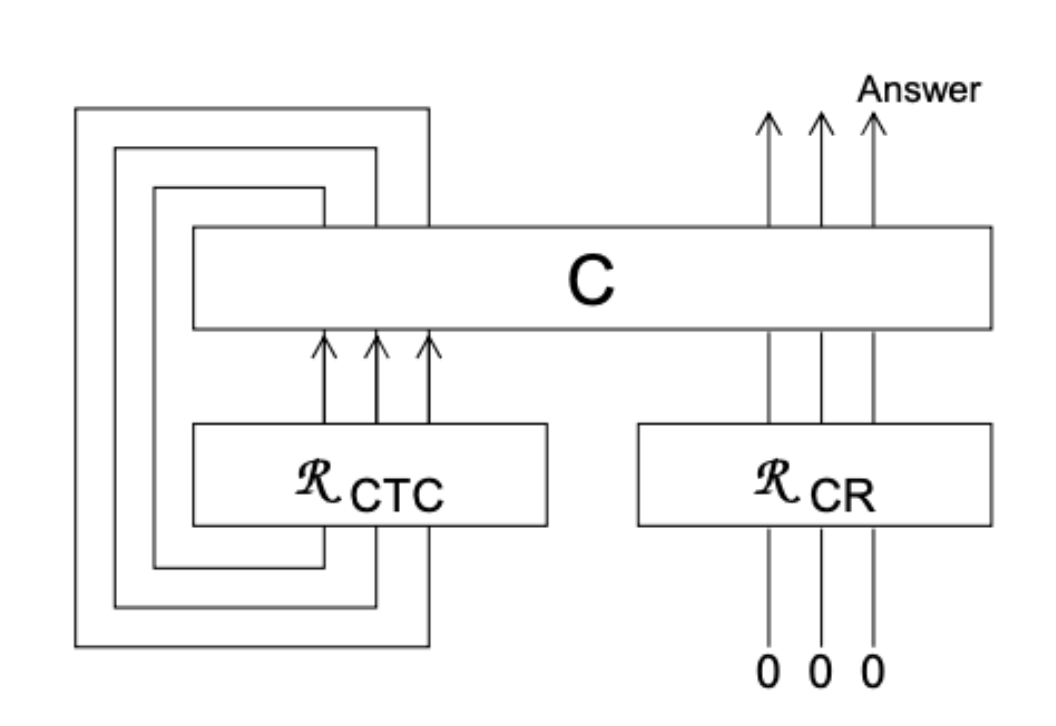}
         \caption{A general scheme \citep{aaronson-quantum}}
         \label{fig2-b}
     \end{subfigure}
        \caption{The $\text{TM}_{\text{CTC}}$ computational model}
        \label{fig2}
\end{figure}

Similar to a classical Turing machine, we assume that both tapes have infinitely many cells, though we can use a finite number of them in each Turing program. Hence we can show the Turing machine’ information with a pair of binary strings $(x, y)$, such that $x$ is the content of $\text{R}_{\text{CR}}$ and $y$ is the content of $\text{R}_{\text{CTC}}$.
\\\
Moreover, for each input $x$ of the machine, there is an infinite dimensional stochastic matrix $S_x$, which maps every binary string $y \in \{ 0, 1 \}^*$ to a binary string $S_x(y) \in \{ 0, 1 \}^*$ with a defined probability. In fact, this stochastic matrix is a \textbf{Markov chain}, and a fixed-point for the operator of each machine is equivalent to a \textbf{stationary distribution} for its corresponding Markov chain.

\subsection{Proof of $\text{TM}_{\text{CTC}}$-Computability of Halting Problem}

\textit{Aaronson} et al. have introduced a $\text{TM}_{\text{CTC}}$ program for solving the halting problem in \citep{aaronson-computability}. This program takes a $\langle P \rangle$, the description of a Turing machine $P$, on $\text{R}_{\text{CR}}$ as input and determines whether $P$ without any inputs will eventually halt or not.
\\\
For doing this, we consider $\sigma_t$ as the configuration of Turing machine $P$ within $t$ steps. Thus the state $\sigma_0$ for $P()$ is obvious. Also, $\sigma_{t + 1}$ is simply attainable from $\sigma_t$ by running one more step of $P$, for every $t$. Additionally, for every arbitrary string $y$, and by knowing $\langle P \rangle$, it is easy to specify whether there is a $t$ for which $y = \sigma_t$ or not.
\\\
Furthermore, we call $\sigma_t$ a \textbf{halting history} if it demonstrates that $P$ halts in the $t^\text{th}$ step of running; otherwise, we name it a \textbf{non-halting history}.
\\\
At this time, the $\text{TM}_{\text{CTC}}$, by taking $\langle P \rangle$ on the $\text{R}_{\text{CR}}$, writes an arbitrary string $y$ on the $\text{R}_{\text{CTC}}$, and we define the function $S_{\langle P \rangle}(y)$ as follows:

\begin{itemize}
    \item[1.] If there existed a $t$ such that $y = \sigma_t$:

    \begin{itemize}
        \item[1.1.] If $\sigma_t$ was a halting history, then $S_{\langle P \rangle}(y) = y$, and it would output $1$ on the $\text{R}_{\text{CR}}$.

        \item[1.2.] Otherwise, $S_{\langle P \rangle}(y) = \sigma_{t + 1}$ with probability $\frac{1}{2}$, and $S_{\langle P \rangle}(y) = \sigma_{0}$ with probability $\frac{1}{2}$.
    \end{itemize}

    \item[2.] Otherwise, $S_{\langle P \rangle}(y) = \sigma_0$.
\end{itemize}

Now, if $P()$ halts, there is a $t$, for which $\sigma_t$ is a halting history, and $y = \sigma_t$ is the only fixed-point of the operator $S_{\langle P \rangle}(y)$. As a result, $\text{TM}_{\text{CTC}}$ will output $1$, meaning that $P()$ will halt.
\\\
In contrast, if $P()$ never halts, the geometric distribution over steps is the only fixed point of the operator $S_{\langle P \rangle}$. In other words, $P(\sigma_i) = (\frac{1}{2})^{i + 1}$, so the $\text{TM}_{\text{CTC}}$ will never halt.

\section{The Problem with the Proof}

In this section, we want to bring up some doubts, as a result of the physical constraints of our universe, about the aforementioned proof.

To begin with, let us study CTC in a physical context.
Having this in mind, consider a universe containing CTCs and therefore, the ability to violate chronology.
Suppose that a particle of it, namely $x$, which is born at the moment $t_0$, goes toward the future\footnote{Actually, $x$ does not move in time; instead, time naturally passes for $x$, and since time-like lines are close in space-time, at some point, by moving in the same direction as the future light cone, the $t$ component decreases for $x$'s coordinate in space-time.} until reaching the moment $t_{-1}$, a time before $x$'s birth.
Then, if $x$ continues its motion on the CTC and reaches the moment $t_0$ without any damages, assuming that all the other particles in the first visit of $t_0$ by $x$ are identical with the second turn, $x$ will be born again.
Thus, there will be two copies of $x$ in the universe. Consequently, both $x$’s move toward the future; so that, when they arrive at the moment $t_0$ and $x$ is reborn, there are three copies of it in the world.
Likewise, this leads to having infinitely many $x$’s in the world which seems impossible in our universe.
Hence, we argue that $x$ must be destroyed at some point between $t_{-1}$ and $t_0$, which will be restated as the \textit{strong axiom} in section 3.1.

\begin{figure}[H]
    \centering
    \includegraphics[width=0.22\textwidth]{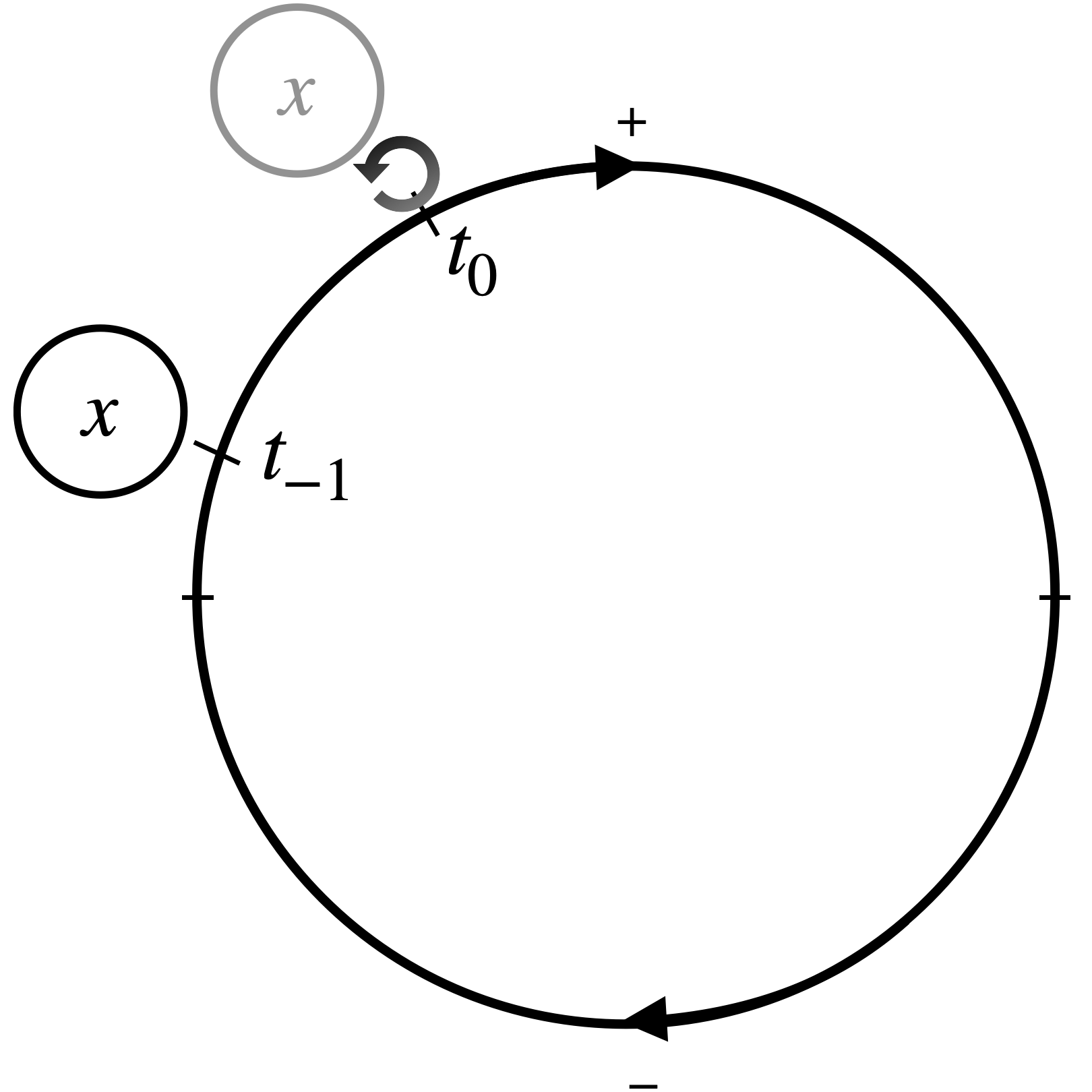}
    \caption{\label{fig4}The motion of particle $x$ on a CTC.}
\end{figure}

The same problem can be discussed for a Turing machine equipped with CTC running the program of the previous section.
Suppose that in our physical world, the $\text{TM}_{\text{CTC}}$ is created and starts its computation at moment $t_0$ with a $\langle P \rangle$ on its $\text{R}_{\text{CR}}$, and an arbitrary $y$ on its $\text{R}_{\text{CTC}}$.
If by the passage of time, the machine returns to the moment $t_0$, it will be recreated and therefore according to the argument we made above, there will be infinitely many of it in the world.
Thus, by reaching a moment $t_{-1} < t_0$, the $\text{TM}_{\text{CTC}}$ should be destroyed before $t_0$, which will be restated as the \textit{weak axiom} in section 3.1, and, as a result, is not able to use any information it has acquired from the future. Meaning that $y$ is always the arbitrary string and will not get any closer to a halting history. More precisely, if we trace $S_{\langle P \rangle}(y)$, we have:
\begin{itemize}
    \item[1.] If there existed a $t$ such that $y = \sigma_t$:
   
    \begin{itemize}
        \item[1.1.] If $\sigma_t$ was a halting history, then $y$ will remain on the $\text{R}_{\text{CTC}}$, and $\text{R}_{\text{CR}}$ would output $1$. Basically, this is the only case that the program responds correctly without any usage of CTCs. However, there is no guarantee that it always happens.

         \item[1.2.] If $\sigma_t$ was a non-halting history, either another step of the Turing machine is supposed to be appended to $y$, or it ought to be changed to the trivial configuration, $\sigma_0$, until the upcoming visit of $t_0$. Nevertheless, by the destruction of the Turing machine before $t_0$, this progress will be lost, and the $\text{TM}_{\text{CTC}}$ will always start its computation with y on the $\text{R}_{\text{CTC}}$.
    \end{itemize}

    \item[2.] Otherwise, the trivial configuration is supposed to be written on the $\text{R}_{\text{CTC}}$ before the subsequent visit of $t_0$, which, similar to the previous case, will not be accessible.
\end{itemize}

Moreover, there are some ambiguities about CTCs in the physical universe that have not been addressed during the proof. For instance, all definitions are about microscopic particles; how are they extended for macroscopic particles such as Turing machines? Further, time is considered just like space, which seems not reasonable.
\section{Our Proposed Solution}

We aim to study the problem brought up in the previous section from two different points of view. Firstly, we try to propose a solution by considering the rules of classical physics over the universe, after which we will look at the problem outside the current world.

\subsection{Solving the Problem in Our Current Universe}

In section 2, we raised a problem, describing that the lasting of particles for a whole round on a CTC results in having an infinite number of each particle in the universe.
It seems like the issue does not match our physical intuition, even though it appears to be mathematically consistent. In other words, we assume that the problem only exists in our physical world, while the mathematical world is a possible and untroubled platform for $\text{TM}_{\text{CTC}}$.
\\\
As a consequence of the argument we made in the previous section, the two following axioms can be stated for a classical universe containing CTCs:
\begin{quote}
    \textbf{Strong axiom}: No particle survives a full round of movement on a CTC and will be destroyed before returning to its starting point in space-time.
    \\\
    \textbf{Weak axiom}: Every Turing machine rounding on a CTC, will be destroyed before returning to its starting point in space-time.
\end{quote}

Now premising at least the \textit{weak axiom}, we will solve the problem by transferring data between Turing machines.
\\\
We should remark that, according to the definitions in the first section, moving in the positive or negative direction of time mean movement toward the future or past, respectively.
\\\
Suppose a $\text{TM}_{\text{CTC}}$, namely $M$, starts its calculation at time $t_0$, moves in positive time until time $t_1$, and since then, continues moving in negative time, reaches $t_2 \leq t_0$ and moves again in positive time to reach $t_1$ again. In this case, according to the \textit{weak axiom}, $M$ will be destroyed at time $t_3$ when $t_2 \leq t_3 \leq t_1$ and $M$ is moving in negative time, or $t_3'$ when $t_2 \leq t_3' \leq t_0$ and $M$ is moving in positive time.
\\\
Now to solve the problem, suppose that another $\text{TM}_{\text{CTC}}$, namely $M'$, can be placed in $M$’s path before reaching $t_3$ or $t'_3$, which we call the \textbf{data transferring hypothesis}. Thus, the calculated output of $M$ until that moment can be used as input of $M'$. Then, $M'$ will move from $t_3$ or $t'_3$ to $t_0$, when it gives its data as input of $M$, without any process or alteration.
\\\
Within this approach, particles and Turing machines will be destroyed during an entire round movement on CTC; nevertheless, data can travel in time and remains stable. Thus, $M$ can use the information gathered by moving on a CTC, and so given claims and proofs of $\text{TM}_{\text{CTC}}$ will be valid.

\begin{figure}[H]
     \centering
     \begin{subfigure}[b]{0.49\textwidth}
         \centering
         \includegraphics[width=0.5\textwidth]{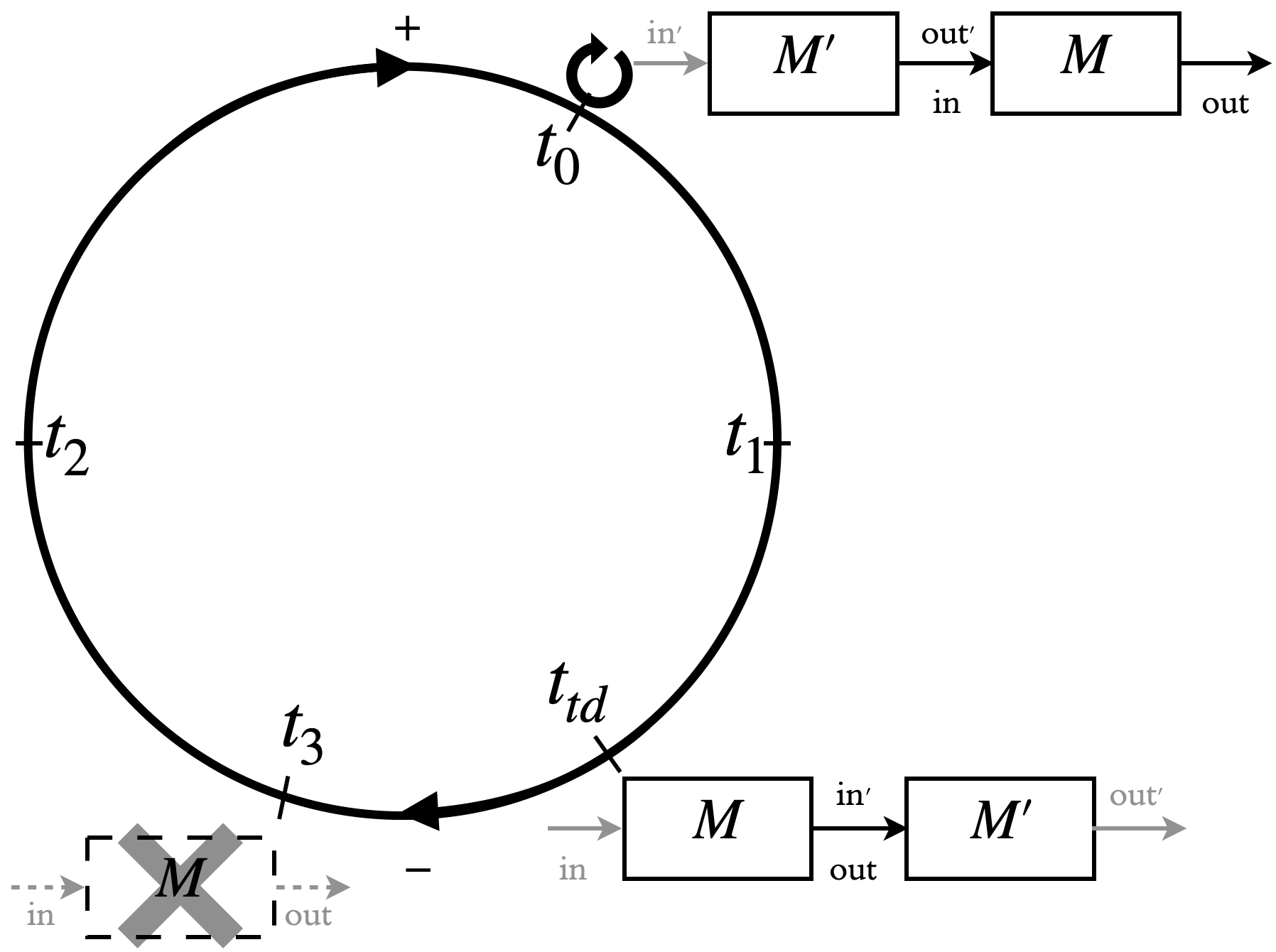}
         \caption{}
         \label{fig5-a}
     \end{subfigure}
     \hfill
     \begin{subfigure}[b]{0.49\textwidth}
         \centering
         \includegraphics[width=0.72\textwidth]{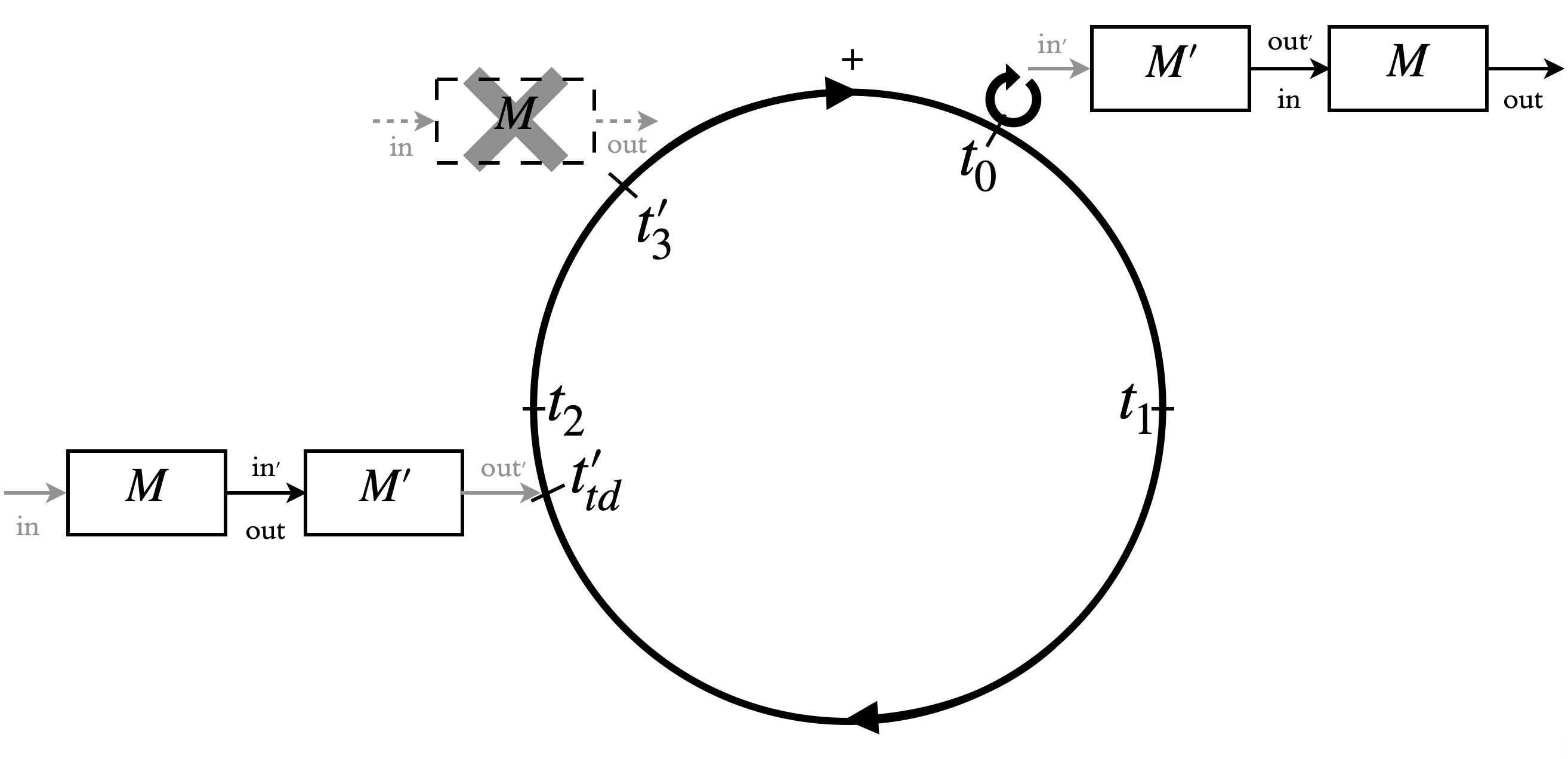}
         \caption{}
         \label{fig5-b}
     \end{subfigure}
        \caption{Data transferring between Turing machines $M$ and $M'$.}
        \label{fig5}
\end{figure}

Hence, in a universe where \textit{data transferring} was possible, as \textit{Aaronson} claimed, not only $\text{P}_{\text{CTC}} = \text{PSAPCE}_{\text{CTC}}$ but the halting problem would be computable.

\subsubsection{Prerequisites for the Data Transferring Hypothesis}

We demonstrate the necessary conditions of the \textit{data transferring hypothesis} via the following conversation:

\begin{quote}
    \textbf{A:} A prerequisite of the \textit{data transferring hypothesis} is the existence of Turing machines throughout time; since otherwise, in the first round of the CTC, there was no Turing machine in space-time until 1936, when Turing machines were invented. However, in the subsequent rounds, there should exist at least one Turing machine at every moment, including before 1936, to establish the \textit{data transferring hypothesis}. This leads to visiting dissimilar universes in different cycles on the CTC.
    
    \begin{figure}[H]
        \centering
        \includegraphics[width=0.9\textwidth]{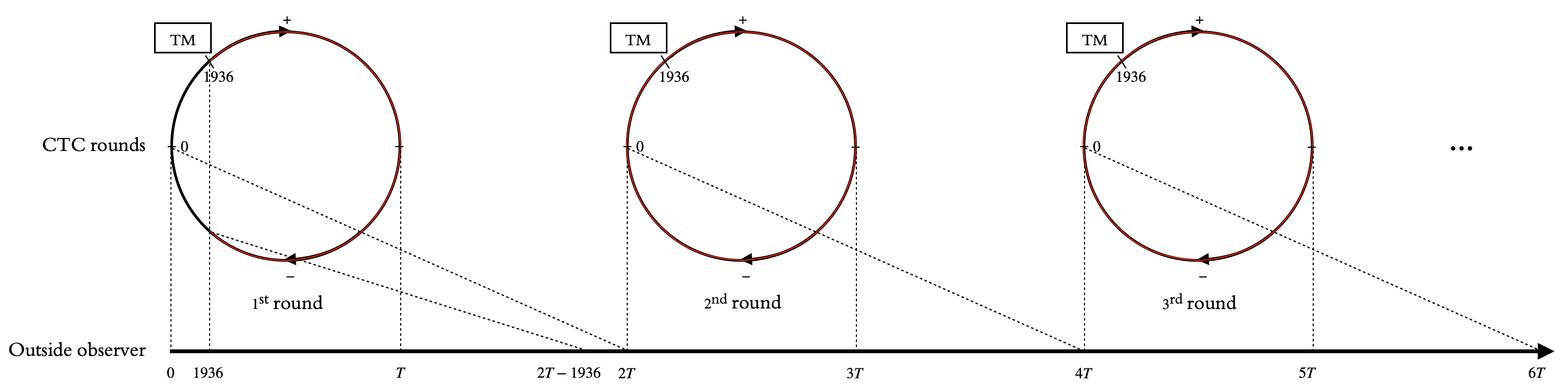}
        \caption{\label{fig6}The different cycles the \textit{data transferring hypothesis} requires. The red lines show the time that Turing machines exist which is not the same in the first and second rounds.}
    \end{figure}

    \textbf{B:} Due to the fact that CTC is a time-like line, is it necessary to be unique, start from the origin of time, and continue to a far future? 
    \\
    For instance, can we assume that every particle is able to own its particular CTC and let the corresponding CTC to a Turing machine be a small loop, passing only 2022? In other words, while the first-ever Turing machine in the world rounds on a CTC and according to the \textit{weak axiom}, is destroyed before reaching 1936, another Turing machine rounds on a CTC passing only 2022, over the entire path of which Turing machines always exist. Moreover, some particles may not round on a CTC.

    \begin{figure}[H]
        \centering
        \includegraphics[width=0.135\textwidth]{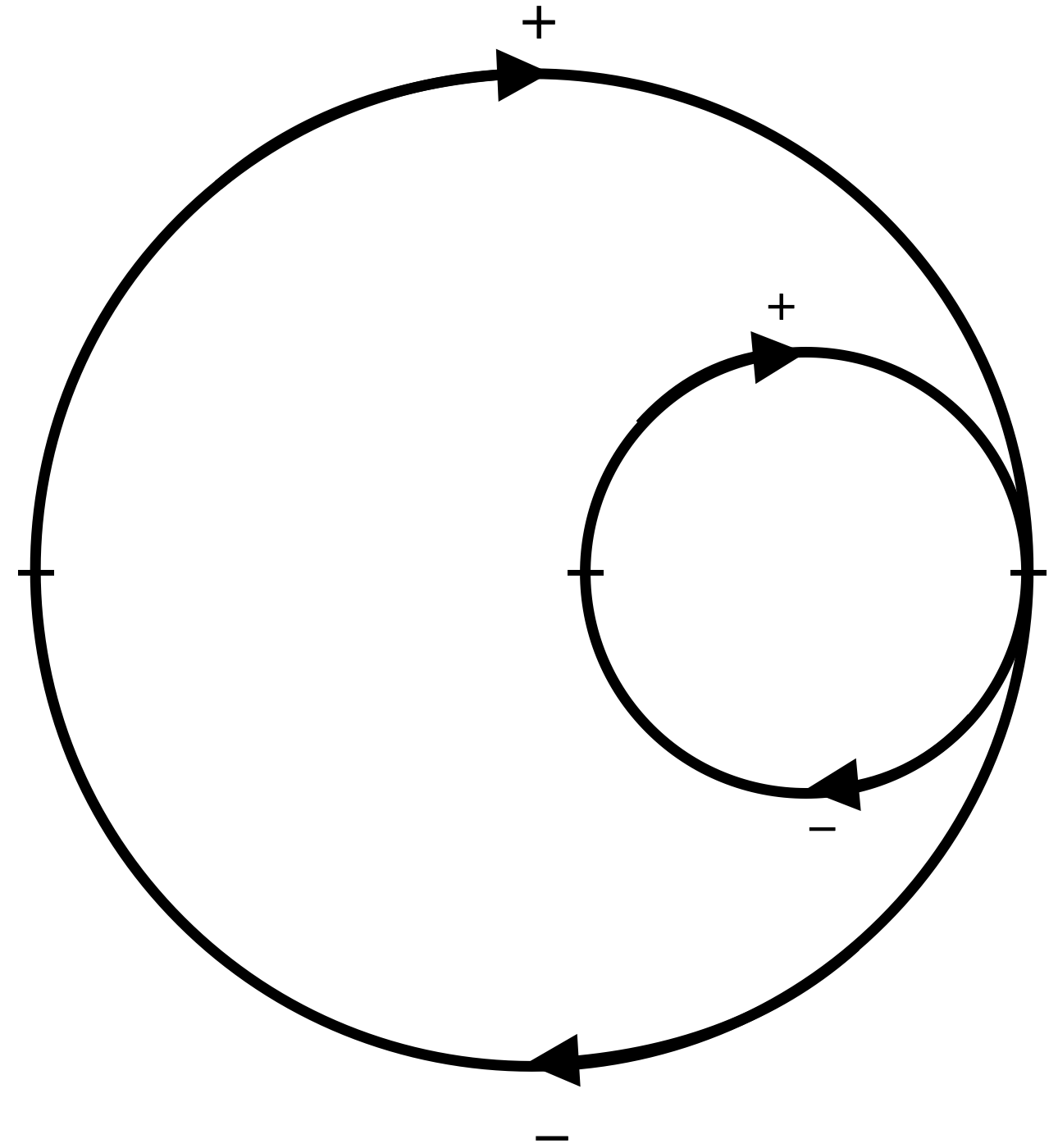}
        \caption{\label{fig7}Various CTCs for different particles.}
    \end{figure}
    
    \textbf{A:} In this case, how might the communication of different particles be?
    \\
    \textbf{B:} We can say that if two particles on a CTC round interacted with each other,  they would also interact on all other CTC rounds. Like \textit{Novikov’s self-consistency principle} which says that only those particles can travel back in time that does not change the past.
    \\
    \textbf{A:} Consider a piece of stone, half of which rounds on a 20 million-year length CTC and the other half rounds on a 2 million-year length CTC. Then what would happen to this stone?
    \\
    Similarly, what if we consider these two CTCs for two separate stones?
    \\
    \textbf{B:} Hence, having a single CTC for all particles in the universe seems reasonable. However, is it necessary for the CTC to return to a moment before 1936, the origin of Turing machines?
    \\
    \textbf{A:} Assuming CTC does not return a specific moment is a strong hypothesis. Therefore, It is reasonable to discuss other models in which the starting and ending points are not necessarily similar.

\end{quote}

\begin{figure}[H]
     \centering
     \begin{subfigure}[b]{0.49\textwidth}
         \centering
         \includegraphics[width=0.35\textwidth]{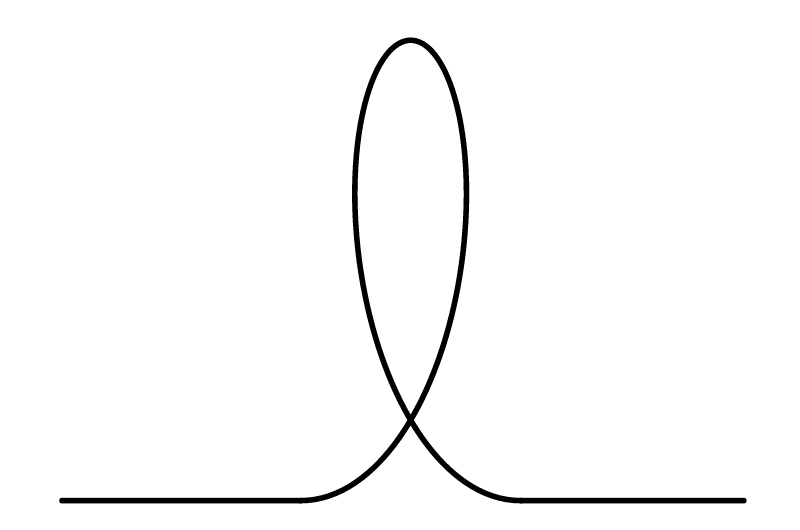}
         \caption{Loop}
         \label{fig8-a}
     \end{subfigure}
     \hfill
     \begin{subfigure}[b]{0.49\textwidth}
         \centering
         \includegraphics[width=0.43\textwidth]{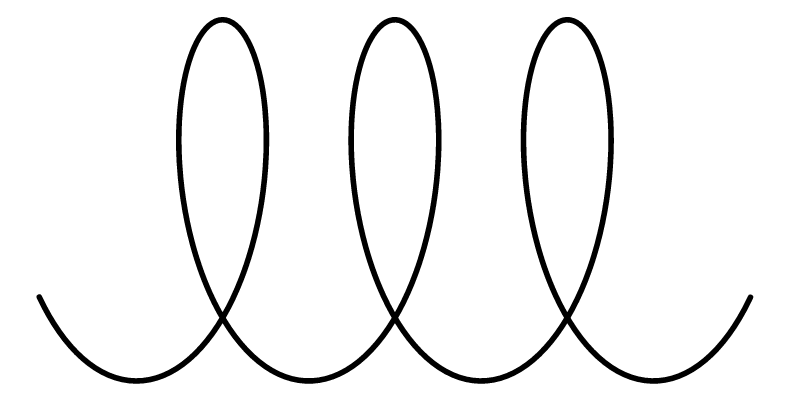}
         \caption{Loop$^\infty$}
         \label{fig8-b}
     \end{subfigure}
        \caption{Chronology-violating models rather than CTC.}
        \label{fig8}
\end{figure}

By the above discussion, the \textit{data transferring hypothesis} also seems like not possible in our universe. So, we should discuss possible situations that a universe containing CTCs may have in the next section.
\subsubsection{Possible Situations}

In this section, we want to discuss almost all possible cases for a universe equipped with CTC. For each of these, we will discuss problems that can appear and try to propose solutions for them. It should be noted that there might be other cases we have missed, and we would be glad if you informed us of any you have reached.
\paragraph{The weak axiom without further conditions}
Assuming there is no concentration on the cycle on which the CTC rounds, nor any specific condition for the strong axiom, we only know that each particle in space-time will be destroyed before returning to its birth moment. Hence, as we discussed in the previous section, the \textit{data transferring hypothesis} needs Turing machines to be existed throughout time, while we know that it is not satisfied in our current universe. Thus, although the \textit{data transferring hypothesis} would be a helpful claim in possible universes, it does not work in our world.

\paragraph{Returning to an approximately equivalent universe}
In this case, suppose that particles moving on a CTC return to a universe approximately equivalent to the one in the previous cycle instead of an identical universe, meaning that some items of the universe can be excepted to be different in two rounds on the CTC; such as the existence of Turing machines. In other words, in the first-ever pass of space-time, Turing machines were invented in 1936, before which there was no Turing machine in the world. However, in all following cycles resulting from CTCs, Turing machines exist throughout time.
\\
It should be remarked that according to the \textit{strong axiom} we inferred from the problem discussed in section 2, no $\text{TM}_{\text{CTC}}$ remains alive in a whole cycle of CTC. Rather, the concept of Turing machines, with necessarily different instances, is well-known at all times.
\\
Thus, the prerequisites for the \textit{data transferring hypothesis} will be held, and therefore, according to what we explained above, the deduced results about complexity and computability classes of $\text{TM}_{\text{CTC}}$ are valid.

\paragraph{The weak axiom in the last possible moment}
Suppose that each particle on a CTC can move safely and freely in positive and negative time directions until it returns just before its birth moment when it is destroyed and immediately recreated. Therefore, not only holds the \textit{strong axiom} and there is no more than one version of any particle in the universe, but from a third person’s point of view, who is out of space-time, all creatures are always alive.
\\\
However, in this case, the problem remains for Turing machines, since by destruction of a Turing machine, all of its information will be missed and will not be accessible after its recreation. Thus, it seems like the Turing machine’s obtained data has been erased and cannot be used in the next cycle on CTC.

\paragraph{Transferring data between various Turing Machines}
Implementing the \textit{data transferring hypothesis} amongst more than two Turing machines would be another situation that comes to mind for solving the problem. However, still requires in every moment of the cycle, at least one Turing machine exists in order to carry the information. Therefore, it does not address the issue, and the problem still remains.

\paragraph{Transferring data between different time directions}
We have already defined that the movement of particles on the CTC in the positive or negative direction of time depends on whether they are moving toward the future or the past. Suppose that the \textit{data transferring hypothesis} is possible between a Turing machine moving in the positive direction and one in the negative direction, meaning that a Turing machine $M_1$, which in moment $t_0$ goes toward the past, can transfer its information to a Turing machine $M_2$, which goes toward the future in moment $t_0$. Hence, the existence of Turing machines at all times is no more required for the \textit{data transferring hypothesis}. However, it should be discussed whether such communication between particles with different time directions is possible. For instance, due to movement in different time directions, the two Turing machines can touch each other just in a second, after which they have no access to each other, and therefore, transferring an arbitrary amount of information in a second must be possible, which seems not.
This case indeed can solve the problem; nevertheless, it is unlikely to be possible.

 \begin{figure}[H]
        \centering
        \includegraphics[width=0.18\textwidth]{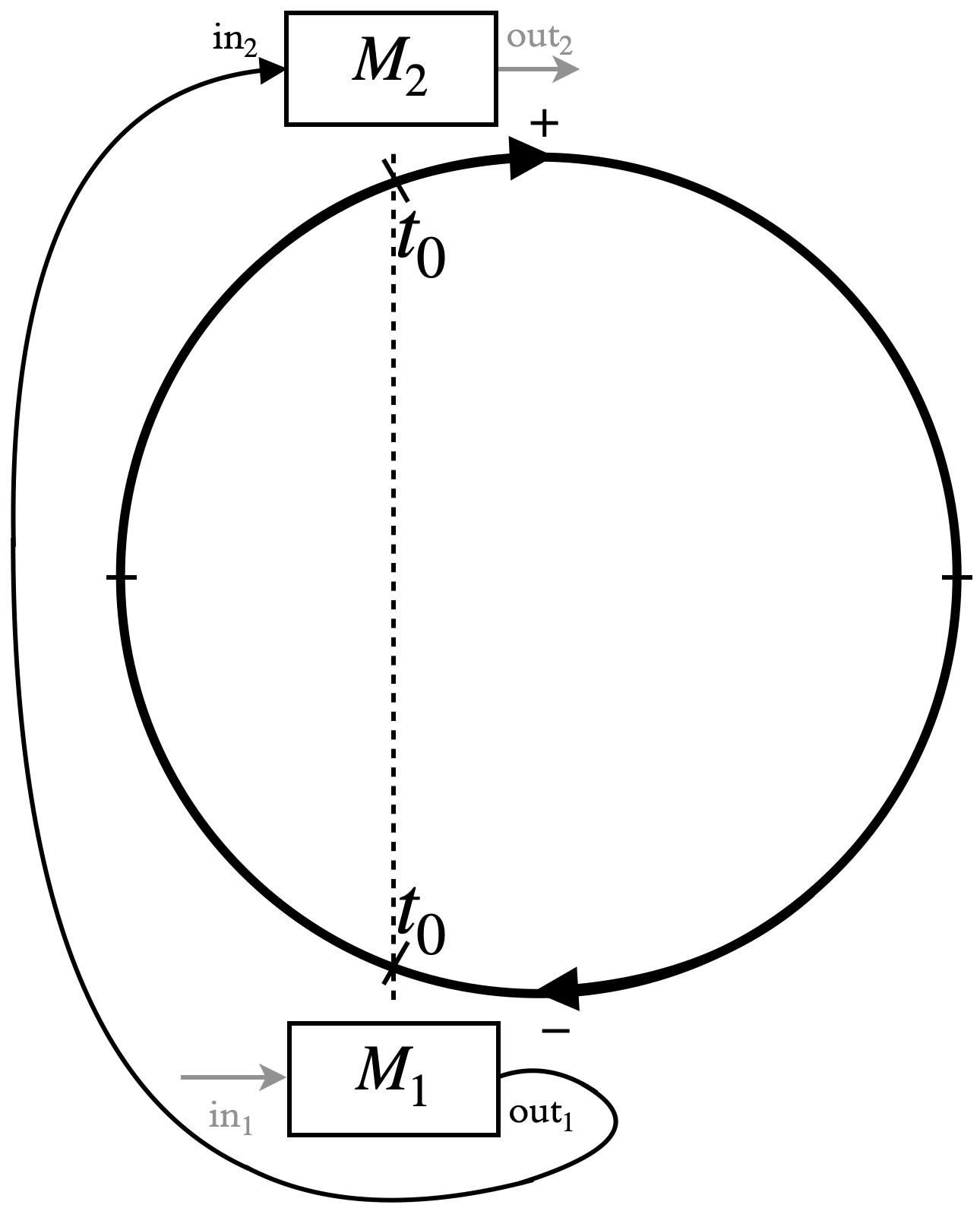}
        \caption{\label{fig9} Data transferring between Turing machines $M_1$, moving to the past, and $M_2$, moving toward the future, in $t_0$.}
    \end{figure}
    
After discussing all cases we came into, we request readers notify the authors if they attained any other reasonable scenario.
\section{Conclusion}

In this paper, after studying the definition of closed time-like curves and the fundamental difference revisiting the past within them has with the initial idea of time travel, we considered the proof of computability of the halting problem by $\text{TM}_\text{CTC}$  provided in \citep{aaronson-computability}. It raised the physical objection discussed in section 2 as the \textit{weak} and \textit{strong axioms}, explaining that no particle of the universe survives a full round on a CTC, leading to the inability of $\text{TM}_\text{CTC}$  to solve the halting problem.
\\
Then, we tried to address this issue using the \textit{data transferring hypothesis}, which basically utilizes another $\text{TM}_\text{CTC}$  as a medium for storing information over a cycle of CTC. The \textit{data transferring hypothesis} also required other seemingly infeasible conditions, such as the existence of Turing machines throughout time.
\\
Finally, we reviewed all the possible physical scenarios as far as we could think of for a universe containing CTCs in the last section, albeit there might be additional scenarios that we welcome being acquainted with.

\bibliographystyle{johd}
\bibliography{bib}

\begin{thebibliography}{}

\bibitem [\protect \citeauthoryear {%
Aaronson%
}{%
Aaronson%
}{%
{\protect \APACyear {2005}}%
}]{%
aaronson-complexity}
\APACinsertmetastar {%
aaronson-complexity}%
\begin{APACrefauthors}%
Aaronson, S.%
\end{APACrefauthors}%
\unskip\
\newblock
\APACrefYearMonthDay{2005}{}{}.
\newblock
{\BBOQ}\APACrefatitle {Guest Column: \uppercase{NP}-complete Problems and
  Physical Reality} {Guest column: \uppercase{NP}-complete problems and
  physical reality}.{\BBCQ}
\newblock
\APACjournalVolNumPages{ACM Sigact News}{36}{1}{30-52}.
\PrintBackRefs{\CurrentBib}

\bibitem [\protect \citeauthoryear {%
Aaronson%
, Bavarian%
\BCBL {}\ \BBA {} Gueltrini%
}{%
Aaronson%
\ \protect \BOthers {.}}{%
{\protect \APACyear {2016}}%
}]{%
aaronson-computability}
\APACinsertmetastar {%
aaronson-computability}%
\begin{APACrefauthors}%
Aaronson, S.%
, Bavarian, M.%
\BCBL {}\ \BBA {} Gueltrini, G.%
\end{APACrefauthors}%
\unskip\
\newblock
\APACrefYearMonthDay{2016}{}{}.
\newblock
{\BBOQ}\APACrefatitle {Computability Theory of Closed Timelike Curves}
  {Computability theory of closed timelike curves}.{\BBCQ}
\newblock
\APACjournalVolNumPages{arXiv preprint arXiv:1609.05507}{}{}{}.
\PrintBackRefs{\CurrentBib}

\bibitem [\protect \citeauthoryear {%
Aaronson%
\ \BBA {} Watrous%
}{%
Aaronson%
\ \BBA {} Watrous%
}{%
{\protect \APACyear {2009}}%
}]{%
aaronson-quantum}
\APACinsertmetastar {%
aaronson-quantum}%
\begin{APACrefauthors}%
Aaronson, S.%
\BCBT {}\ \BBA {} Watrous, J.%
\end{APACrefauthors}%
\unskip\
\newblock
\APACrefYearMonthDay{2009}{}{}.
\newblock
{\BBOQ}\APACrefatitle {Closed timelike curves make quantum and classical
  computing equivalent} {Closed timelike curves make quantum and classical
  computing equivalent}.{\BBCQ}
\newblock
\APACjournalVolNumPages{Proceedings of the Royal Society A: Mathematical,
  Physical and Engineering Sciences}{465}{2102}{631-647}.
\PrintBackRefs{\CurrentBib}

\bibitem [\protect \citeauthoryear {%
Brennan%
}{%
Brennan%
}{%
{\protect \APACyear {1997}}%
}]{%
brennan-hitler}
\APACinsertmetastar {%
brennan-hitler}%
\begin{APACrefauthors}%
Brennan, J\BPBI H.%
\end{APACrefauthors}%
\unskip\
\newblock
\APACrefYear{1997}.
\newblock
\APACrefbtitle {Time Travel: A New Perspective} {Time travel: A new
  perspective}.
\newblock
\APACaddressPublisher{}{Llewellyn Publications}.
\PrintBackRefs{\CurrentBib}

\bibitem [\protect \citeauthoryear {%
Craig%
}{%
Craig%
}{%
{\protect \APACyear {1988}}%
}]{%
craig-newcomb}
\APACinsertmetastar {%
craig-newcomb}%
\begin{APACrefauthors}%
Craig, W\BPBI L.%
\end{APACrefauthors}%
\unskip\
\newblock
\APACrefYearMonthDay{1988}{}{}.
\newblock
{\BBOQ}\APACrefatitle {Tachyons, Time Travel, and Divine Omniscience}
  {Tachyons, time travel, and divine omniscience}.{\BBCQ}
\newblock
\APACjournalVolNumPages{The Journal of Philosophy}{85}{3}{135-150}.
\PrintBackRefs{\CurrentBib}

\bibitem [\protect \citeauthoryear {%
Deutsch%
}{%
Deutsch%
}{%
{\protect \APACyear {1991}}%
}]{%
deutsch-ctc}
\APACinsertmetastar {%
deutsch-ctc}%
\begin{APACrefauthors}%
Deutsch, D.%
\end{APACrefauthors}%
\unskip\
\newblock
\APACrefYearMonthDay{1991}{}{}.
\newblock
{\BBOQ}\APACrefatitle {Quantum mechanics near closed timelike lines} {Quantum
  mechanics near closed timelike lines}.{\BBCQ}
\newblock
\APACjournalVolNumPages{Physical Review D}{44}{10}{3197–3218}.
\PrintBackRefs{\CurrentBib}

\bibitem [\protect \citeauthoryear {%
Einstein%
}{%
Einstein%
}{%
{\protect \APACyear {1905}}%
}]{%
einstein-special-relativity}
\APACinsertmetastar {%
einstein-special-relativity}%
\begin{APACrefauthors}%
Einstein, A.%
\end{APACrefauthors}%
\unskip\
\newblock
\APACrefYearMonthDay{1905}{}{}.
\newblock
{\BBOQ}\APACrefatitle {On the Electrodynamics of Moving Bodies} {On the
  electrodynamics of moving bodies}.{\BBCQ}
\newblock
\APACjournalVolNumPages{Annalen der physik}{17}{10}{891-921}.
\PrintBackRefs{\CurrentBib}

\bibitem [\protect \citeauthoryear {%
Einstein%
}{%
Einstein%
}{%
{\protect \APACyear {1916}}%
}]{%
einstein-general-relativity}
\APACinsertmetastar {%
einstein-general-relativity}%
\begin{APACrefauthors}%
Einstein, A.%
\end{APACrefauthors}%
\unskip\
\newblock
\APACrefYearMonthDay{1916}{}{}.
\newblock
{\BBOQ}\APACrefatitle {The Foundation of the General Theory of Relativity} {The
  foundation of the general theory of relativity}.{\BBCQ}
\newblock
\APACjournalVolNumPages{Annalen der physik}{49}{}{769-822}.
\PrintBackRefs{\CurrentBib}

\bibitem [\protect \citeauthoryear {%
Friedman%
\ \protect \BOthers {.}}{%
Friedman%
\ \protect \BOthers {.}}{%
{\protect \APACyear {1990}}%
}]{%
novikov-consistency}
\APACinsertmetastar {%
novikov-consistency}%
\begin{APACrefauthors}%
Friedman, J.%
, Morris, M\BPBI S.%
, Novikov, I\BPBI D.%
, Echeverria, F.%
, Klinkhammer, G.%
, Thorne, K\BPBI S.%
\BCBL {}\ \BBA {} Yurtsever, U.%
\end{APACrefauthors}%
\unskip\
\newblock
\APACrefYearMonthDay{1990}{}{}.
\newblock
{\BBOQ}\APACrefatitle {Cauchy problem in spacetimes with closed timelike
  curves} {Cauchy problem in spacetimes with closed timelike curves}.{\BBCQ}
\newblock
\APACjournalVolNumPages{Physical Review D}{42}{6}{1915}.
\PrintBackRefs{\CurrentBib}

\bibitem [\protect \citeauthoryear {%
Gödel%
}{%
Gödel%
}{%
{\protect \APACyear {1949}}%
}]{%
godel-ctc}
\APACinsertmetastar {%
godel-ctc}%
\begin{APACrefauthors}%
Gödel, K.%
\end{APACrefauthors}%
\unskip\
\newblock
\APACrefYearMonthDay{1949}{}{}.
\newblock
{\BBOQ}\APACrefatitle {An Example of a New Type of Cosmological Solutions of
  Einstein’s Field Equations of Gravitation} {An example of a new type of
  cosmological solutions of einstein’s field equations of
  gravitation}.{\BBCQ}
\newblock
\APACjournalVolNumPages{Reviews of Modern Physics}{21}{3}{447–450}.
\PrintBackRefs{\CurrentBib}

\bibitem [\protect \citeauthoryear {%
Hawking%
}{%
Hawking%
}{%
{\protect \APACyear {1999}}%
}]{%
hawking-time-travel}
\APACinsertmetastar {%
hawking-time-travel}%
\begin{APACrefauthors}%
Hawking, S\BPBI W.%
\end{APACrefauthors}%
\unskip\
\newblock
\APACrefYearMonthDay{1999}{}{}.
\newblock
{\BBOQ}\APACrefatitle {Space and Time Warps (Public Lecture)} {Space and time
  warps (public lecture)}.{\BBCQ}
\newblock
\APACaddressPublisher{\url{https://www.hawking.org.uk/in-words/lectures/space-and-time-warps}}{The
  Stephen Hawking Estate}.
\PrintBackRefs{\CurrentBib}

\bibitem [\protect \citeauthoryear {%
Hawking%
\ \BBA {} Ellis%
}{%
Hawking%
\ \BBA {} Ellis%
}{%
{\protect \APACyear {1973}}%
}]{%
hawking-grandfather}
\APACinsertmetastar {%
hawking-grandfather}%
\begin{APACrefauthors}%
Hawking, S\BPBI W.%
\BCBT {}\ \BBA {} Ellis, G\BPBI F.%
\end{APACrefauthors}%
\unskip\
\newblock
\APACrefYear{1973}.
\newblock
\APACrefbtitle {The large scale structure of space-time} {The large scale
  structure of space-time}.
\newblock
\APACaddressPublisher{Cambridge, England}{Cambridge University Press}.
\PrintBackRefs{\CurrentBib}

\bibitem [\protect \citeauthoryear {%
Jones%
}{%
Jones%
}{%
{\protect \APACyear {1985}}%
}]{%
jones-fermi}
\APACinsertmetastar {%
jones-fermi}%
\begin{APACrefauthors}%
Jones, E\BPBI M.%
\end{APACrefauthors}%
\unskip\
\newblock
\APACrefYearMonthDay{1985}{}{}.
\newblock
{\BBOQ}\APACrefatitle {"Where is Everybody?" An Account of Fermi's Question}
  {"where is everybody?" an account of fermi's question}.{\BBCQ}
\newblock
\APACjournalVolNumPages{Los Alamos National Lab., NM (USA)}{}{}{}.
\PrintBackRefs{\CurrentBib}

\bibitem [\protect \citeauthoryear {%
Lobo%
\ \BBA {} Crawford%
}{%
Lobo%
\ \BBA {} Crawford%
}{%
{\protect \APACyear {2003}}%
}]{%
lobo-causality}
\APACinsertmetastar {%
lobo-causality}%
\begin{APACrefauthors}%
Lobo, F.%
\BCBT {}\ \BBA {} Crawford, P.%
\end{APACrefauthors}%
\unskip\
\newblock
\APACrefYearMonthDay{2003}{}{}.
\newblock
{\BBOQ}\APACrefatitle {Time, closed timelike curves and causality} {Time,
  closed timelike curves and causality}.{\BBCQ}
\newblock
\BIn{} \APACrefbtitle {The Nature of Time: Geometry, Physics and Perception}
  {The nature of time: Geometry, physics and perception}\ (\BPG~289-296).
\newblock
\APACaddressPublisher{}{Springer}.
\PrintBackRefs{\CurrentBib}

\bibitem [\protect \citeauthoryear {%
Nozick%
}{%
Nozick%
}{%
{\protect \APACyear {1969}}%
}]{%
nozick-newcomb}
\APACinsertmetastar {%
nozick-newcomb}%
\begin{APACrefauthors}%
Nozick, R.%
\end{APACrefauthors}%
\unskip\
\newblock
\APACrefYearMonthDay{1969}{}{}.
\newblock
{\BBOQ}\APACrefatitle {Newcomb's Problem and Two Principles of Choice}
  {Newcomb's problem and two principles of choice}.{\BBCQ}
\newblock
\BIn{} N.~Rescher\ (\BED), \APACrefbtitle {Essays in Honor of Carl G. Hempel: A
  Tribute on the Occasion of his Sixty-Fifth Birthday} {Essays in honor of carl
  g. hempel: A tribute on the occasion of his sixty-fifth birthday}\ (\BVOL~24,
  \BPG~114–146).
\newblock
\APACaddressPublisher{Dordrecht}{Springer Netherlands}.
\PrintBackRefs{\CurrentBib}

\bibitem [\protect \citeauthoryear {%
Smith%
}{%
Smith%
}{%
{\protect \APACyear {2021}}%
}]{%
sep-time-travel}
\APACinsertmetastar {%
sep-time-travel}%
\begin{APACrefauthors}%
Smith, N\BPBI J.%
\end{APACrefauthors}%
\unskip\
\newblock
\APACrefYearMonthDay{2021}{}{}.
\newblock
{\BBOQ}\APACrefatitle {Time Travel} {Time travel}.{\BBCQ}
\newblock
\BIn{} E\BPBI N.~Zalta\ (\BED), \APACrefbtitle {The Stanford Encyclopedia of
  Philosophy} {The stanford encyclopedia of philosophy}\ (\PrintOrdinal{Fall
  2021}\ \BEd).
\newblock
\APACaddressPublisher{}{Metaphysics Research Lab, Stanford University}.
\newblock
\APAChowpublished
  {\url{https://plato.stanford.edu/archives/fall2021/entries/time-travel/}}.
\PrintBackRefs{\CurrentBib}

\bibitem [\protect \citeauthoryear {%
van Stockum%
}{%
van Stockum%
}{%
{\protect \APACyear {1938}}%
}]{%
van-ctc}
\APACinsertmetastar {%
van-ctc}%
\begin{APACrefauthors}%
van Stockum, W\BPBI J.%
\end{APACrefauthors}%
\unskip\
\newblock
\APACrefYearMonthDay{1938}{}{}.
\newblock
{\BBOQ}\APACrefatitle {IX.—The gravitational field of a distribution of
  particles rotating about an axis of symmetry} {Ix.—the gravitational field
  of a distribution of particles rotating about an axis of symmetry}.{\BBCQ}
\newblock
\APACjournalVolNumPages{Proceedings of the Royal Society of
  Edinburgh}{57}{}{135-154}.
\PrintBackRefs{\CurrentBib}

\bibitem [\protect \citeauthoryear {%
Wolpert%
\ \BBA {} Benford%
}{%
Wolpert%
\ \BBA {} Benford%
}{%
{\protect \APACyear {2013}}%
}]{%
wolpert-newcomb}
\APACinsertmetastar {%
wolpert-newcomb}%
\begin{APACrefauthors}%
Wolpert, D\BPBI H.%
\BCBT {}\ \BBA {} Benford, G.%
\end{APACrefauthors}%
\unskip\
\newblock
\APACrefYearMonthDay{2013}{}{}.
\newblock
{\BBOQ}\APACrefatitle {The lesson of Newcomb’s paradox} {The lesson of
  newcomb’s paradox}.{\BBCQ}
\newblock
\APACjournalVolNumPages{Synthese}{190}{9}{1637–1646}.
\PrintBackRefs{\CurrentBib}

\end{thebibliography}

\end{document}